# Adiabatic Flame Temperatures for Oxy-Methane, Oxy-Hydrogen, Air-Methane, and Air-Hydrogen Stoichiometric Combustion using the NASA CEARUN Tool, GRI-Mech 3.0 Reaction Mechanism, and Cantera Python Package

**Osama A. Marzouk**

College or Engineering, University of Buraimi, Oman
osama.m@uob.edu.om (corresponding author)

## ABSTRACT

The Adiabatic Flame Temperature (AFT) in combustion represents the maximum attainable temperature at which the chemical energy in the reactant fuel is converted into sensible heat in combustion products without heat loss. AFT depends on the fuel, oxidizer, and chemical composition of the products. Computing AFT requires solving either a nonlinear equation or a larger minimization problem. This study obtained the AFTs for oxy-methane (methane and oxygen), oxy-hydrogen (hydrogen and oxygen), air-methane (methane and air), and air-hydrogen (hydrogen and air) for stoichiometric conditions. The reactant temperature was 298.15 K (25°C), and the pressure was kept constant at 1 atm. Two reaction mechanisms were attempted: a global single-step irreversible reaction for complete combustion and the GRI-Mech 3.0 elementary mechanism (53 species, 325 steps) for chemical equilibrium with its associated thermodynamic data. NASA CEARUN was the main modeling tool used. Two other tools were used for benchmarking: an Excel and a Cantera-Python implementation of GRI-Mech 3.0. The results showed that the AFTs for oxy-methane were 5,166.47 K (complete combustion) and 3,050.12 K (chemical equilibrium), and dropped to 2,326.35 K and 2,224.25 K for air-methane, respectively. The AFTs for oxy-hydrogen were 4,930.56 K (complete combustion) and 3,074.51 K (chemical equilibrium), and dropped to 2,520.33 K and 2,378.62 K for air-hydrogen, respectively. For eight combustion modeling cases, the relative deviation between the AFTs predicted by CEARUN and GRI-Mech 3.0 ranged from 0.064% to 3.503%.

## I. INTRODUCTION

Combustion is a special case of an exothermic chemical reaction in which a fuel reacts with an oxidizer, and heat is released at a high rate. Combustion is an important step in fuel-fired thermal power plants [1], cars with internal combustion engines [2], fuel-fired heating systems, and even domestic devices such as cookers with burners or gas-fired water heaters. One of the characteristic features of combustion is the Adiabatic Flame Temperature (AFT), which is a theoretical upper limit for the combustion products under the assumption of no heat loss, such as loss due to convective heat transfer with the surroundings. The reaction is treated as occurring at a constant enthalpy and a constant pressure, which this study selected to be the common absolute reference value of 1 atm. The experimental measurement of AFT is a very inconvenient and expensive process [3]. Estimating AFT using computational modeling is a powerful alternative, not just because of the eliminated physical setup and saved time, but also because it allows precise control of various combustion parameters, exploration of different combustion situations, and studying their effect on AFT, as well as on the chemical composition of the mixture of combustion products.

Although the term Calculated Adiabatic Flame Temperature (CAFT) [4-5] can be used to specifically refer to AFT estimated numerically rather than obtained experimentally, this study simply uses the term AFT as a calculated value or as a concept in combustion science.

There is a linear relationship between AFT and nitrogen oxide (NOx) emissions. NOx refers to both nitric oxide (NO)

and nitrogen dioxide ($NO_2$) [6], which are air pollutants [7]. Thus, a lower AFT is desirable to limit the harmful air pollutant. This reduction in AFT can be achieved by altering the fuel and blending it with oxygenated fuels, such as methanol or methyl tertiary-butyl ether [8].

High AFT that leads to $NO_x$ emissions can also be avoided by diluting combustion by introducing an excess oxidizer [9] or by Exhaust Gas Recirculation (EGR). Combustion dilution with an EGR rate of 1% ($CO_2$ and $N_2$ gases) decreased AFT and, thus, reduced $NO_x$ emissions by approximately 4.98% [10]. The EGR rate is the mass fraction of exhaust gas in the intake mixture of fresh air and exhaust gas that goes into the engine [11]. In [12], four different EGR rates were investigated for combustion in an internal combustion engine with blended biodiesel fuel. Analyzing several engine performance variables, such as engine cylinder temperature and $NO_x$, CO, hydrocarbon (HC), and soot emissions, it was found that at 100% load, $NO_x$ reduction was 49.9%, 64.3%, and 73.9% for EGR rates of 5%, 10%, and 15%, respectively. In [13], it was shown that flame temperature is the essential cause of $NO_x$ emissions in gas turbines or premixed-flame combustors because the production of thermal $NO_x$ is directly related to temperature. In [14], AFT and the Lower Explosion Limit (LEL) were used to estimate the Limiting Oxygen Concentration (LOC) of fuel-air-inert premixed systems, using nitrogen as an inert gas. If the combustion is made free from nitrogen, either in the fuel or in the oxidizer, the problem of $NO_x$ formation at high temperatures is reduced [15]. There are some applications where an elevated combustion temperature and, thus, a high AFT is desirable, such as oxy-acetylene gas welding [16] and magnetohydrodynamic (MHD) power extraction [17-18], where the fluid possesses electric conductivity [19] that is boosted at high temperatures due to partial ionization.

The aim of low or high AFT can be an important characteristic of a combustion process, which is modeled as a single or a few global chemical reactions, or a larger set of reactions through a detailed reaction mechanism. Proper numerical prediction of AFT requires a robust tool with sufficient information about the species involved in the reaction, as well as the reactions to be enabled during the modeled combustion process. This study focused on the combustion of two gaseous fuels: methane ($CH_4$) and molecular hydrogen ($H_2$), applying the free online tool CEARUN [20], which is a web-based version of the CEA software. CEA performs chemical equilibrium calculations as an optimization problem where the free energy is minimized to find the concentrations of the resultant species that correspond to an arbitrary set of reactant species [21-22]. CEA has a long history that dates back to 1962 [23]. Compared to the equilibrium constants approach to describe chemical equilibrium, the minimization of free energy approach, such as the Gibbs free energy, with a constraint on mass balance [24], as used in CEA, has the advantage of not requiring the set of reactions to be specified first. CEARUN has a large database of 166 species.

For each of the two gaseous fuels considered, methane and hydrogen, AFT is calculated for two combustion modes: oxy-fuel combustion, where the oxidizer is pure molecular oxygen ($O_2$), and air-combustion, where the oxidizer is air. Air is conveniently made available in CEARUN as a single synthetic dry oxidizer species. However, internally it represents a mixture of four gaseous species with the following percentages by mole/volume [25]:

- Molecular nitrogen ($N_2$): 78.0840%

- Molecular oxygen ($O_2$): 20.9476%

- Argon (Ar): 0.9365%

- Carbon dioxide ($CO_2$): 0.0319%

This is a detailed four-component representation, accounting for even the marginal species of $CO_2$. There are studies with simpler three-component representations of air as an $N_2/O_2/Ar$ mixture with mole/volume percentages of 78/21/1% [26] or even just two-component representations as an $N_2/O_2$ mixture with mole/volume percentages of 78/22% [27].

For each of the four cases, two fuels and two oxidizers per fuel, two values were calculated for AFT depending on the combustion mechanism of how the composition of the combustion product mixture is obtained. In the first type of combustion mechanism, the simplest reaction mechanism was assumed that had a single (one-step) global irreversible reaction, corresponding to complete combustion, where all of the fuel and oxygen react together and none of them appear in the products. In the second type of combustion mechanism, small amounts of reactants and new intermediate species, such as carbon monoxide (CO), are allowed according to what a chemical equilibrium state allows, with reversible reactions enabled. The AFT in this detailed combustion mechanism is more realistic and should be lower than its approximate value when assuming irreversible single-step combustion. Calculating both AFTs and comparing them gives a measure of the impact of intermediate reaction steps and the level of error caused by the use of the simplest global reaction mechanism.

This study calculated a total of eight AFTs using the CEARUN tool and eight additional AFTs as benchmarking values for equivalent conditions using the third version of the detailed reaction mechanism of the Gas Research Institute, GRI-Mech 3.0. This reaction mechanism allows the inclusion of up to 53 species, with 325 reactions (reaction steps) [28]. Therefore, this study presents a total of 16 AFTs, of which eight principal values were computed using CEARUN and eight benchmarking values using GRI-Mech 3.0.

Of the eight benchmarking cases, four correspond to a simple one-step reaction with complete combustion. In these cases, AFTs were estimated by solving a nonlinear equation using the Microsoft Excel Goal Seek solver [29-30], combined with tabulated data that incorporate several embedded formulas. The thermodynamic coefficients for a polynomial fitting function of the normalized molar enthalpy needed for this calculation were based on GRI-Mech 3.0, with a small change in the coefficient for $N_2$ to enforce zero enthalpies at the reference temperature of 298.15 K. The low-temperature range coefficient $a_{6L}$ of $N_2$ was changed here from the original value of -1,020.8999 to a new value of -1,021.07188. The air oxidizer was represented as a three-component mixture of $N_2/O_2/Ar$ with mole/volume percentages of 78/21/1%.

For the other four benchmarking cases, GRI-Mech 3.0 was used through the open-source Cantera library for chemical and thermal applications [31-32]. In this study, Cantera was used with Python to infer the chemical composition and simultaneously the AFT. Unlike CEARUN, GRI-Mech 3.0 does not have a customized species that acts as air. For the modeling cases where air was the oxidizer, the air was represented as a three-component mixture of $N_2/O_2/Ar$ with mole/volume percentages of 78/21/1%. The Cantera version used was 2.6.0 and the version of the Python programming environment, as a host for Cantera, was 3.8.8.

## II. MATHEMATICAL FORMULAS FOR GRI-MECH 3.0 (EXCEL SPREADSHEET METHOD)

This section describes how AFT is computed directly using the nonlinear solver tool Goal Seek in MS Excel, using thermodynamic coefficients from GRI-Mech 3.0 for seven species included in modeling complete combustion. AFT is the single unknown variable that satisfies the following condition as a balance of energy under constant overall molar enthalpy ($H$), during the chemical reaction(s) of the combustion process. The pressure was also kept constant as a second condition to allow reaching a unique solution.

$$\sum_i^{Reactants} n_i H_i(T_{ini,i}) = \sum_j^{Products} n_j H_j(T_{ad}) \qquad (1)$$

where $n_i$ is the number of moles of the $i$-th species in the reactants mixture, $n_j$ is the number of moles of the $j$-th species in the products mixture, $H_i(T_{ini,i})$ is the molar enthalpy of the $i$-th species evaluated at its initial temperature $T_{ini,i}$, and $H_j(T_{ad})$ is the molar enthalpy of the $j$-th species evaluated at the common final AFT ($T_{ad}$) for the product species. In this study, the initial temperature of all reactant species was 298.15 K, which is the reference temperature in CEA, where 50 species, such as Ar, $H_2$, $N_2$, and $O_2$, have exactly zero enthalpy at that particular temperature [33].

In GRI-Mech 3.0, the molar enthalpy in a normalized (non-dimensional) form is represented as a fifth-order (6-term) polynomial in the absolute temperature $T$ in Kelvins [34-35].

$$\frac{H}{\Re T} = \begin{cases} a_{1L} + a_{2L}\frac{T}{2} + a_{3L}\frac{T^2}{3} + a_{4L}\frac{T^3}{4} + a_{5L}\frac{T^4}{5} + \frac{a_{6L}}{T} \\ \qquad\qquad T_{min} \leq T \leq T_{mid} \\ a_{1H} + a_{2H}\frac{T}{2} + a_{3H}\frac{T^2}{3} + a_{4H}\frac{T^3}{4} + a_{5H}\frac{T^4}{5} + \frac{a_{6H}}{T} \\ \qquad\qquad T_{mid} \leq T \leq T_{max} \end{cases} \quad (2)$$

where symbol $\Re$ represents the universal (molar) gas constant, the polynomial coefficients $a_{1L}$ - $a_{6L}$ correspond to a low-temperature subrange for the polynomial from $T_{min}$ to $T_{mid}$, and the polynomial coefficients $a_{1H}$ - $a_{6H}$ correspond to a high-temperature subrange for the polynomial from $T_{mid}$ to $T_{max}$. Therefore, each species in GRI-Mech 3.0 is assigned 15 parameters, 12 polynomial coefficients, and 3 boundary temperatures to describe its molar enthalpy as a function of absolute temperature expressed in Kelvins. For most species in GRI-Mech 3.0 (50 out of the total 53 species), the intermediate temperature $T_{mid}$ is 1,000 K. Exceptionally, it is 1,382 K for HCNO (fulminic acid), 1,368 K for HOCN (cyanic acid), and 1,478 K for HNCO (isocyanic acid).

The polynomial in (2) was not introduced by RI-Mech 3.0 itself but was developed earlier by NASA with two more coefficients, $a_{7L}$ and $a_{7H}$ for the entropy in the low- and the high-temperature subranges, respectively. Therefore, this polynomial pattern is one part of three polynomials that are referred to as NASA 7-coefficient polynomials (one more polynomial is used for the molar specific heat capacity at constant pressure $C_p$) regardless of whether the coefficients' values and corresponding temperature limits are provided by NASA itself or not. The GRI-Mech 3.0 coefficients and temperature subranges are generally different from those published by NASA, with a universal gas constant value of 8.314510 J/mol.K [36], although the variation with temperature is similar for both sets of parameters.

The modeled simplified single-step reaction for oxy-fuel complete combustion of methane is:

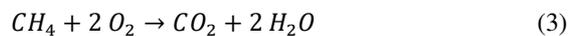

$$CH_4 + 2\,O_2 \rightarrow CO_2 + 2\,H_2O \qquad (3)$$

The modeled simplified single-step reaction for oxy-fuel complete combustion of hydrogen is:

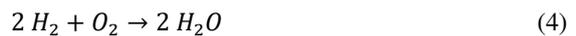

$$2\,H_2 + O_2 \rightarrow 2\,H_2O \qquad (4)$$

The modeled simplified single-step reaction for air-fuel complete combustion of methane is:

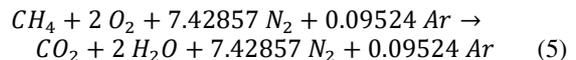

$$CH_4 + 2\,O_2 + 7.42857\,N_2 + 0.09524\,Ar \rightarrow \\ CO_2 + 2\,H_2O + 7.42857\,N_2 + 0.09524\,Ar \qquad (5)$$

The modeled simplified single-step reaction for air-fuel complete combustion of hydrogen is:

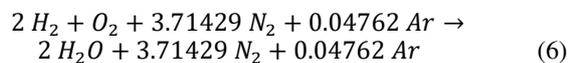

$$2\,H_2 + O_2 + 3.71429\,N_2 + 0.04762\,Ar \rightarrow \\ 2\,H_2O + 3.71429\,N_2 + 0.04762\,Ar \qquad (6)$$

The procedure of solving for AFT requires solving a nonlinear equation in a single unknown. The pressure value is irrelevant in this procedure and does not impact the estimated AFT. This corresponds to an ideal gas assumption, where the molar enthalpy of a gaseous species is a function of its temperature only and is independent of pressure [37].

## III. RESULTS

Figures 1-7 show the visual illustration of the normalized polynomial functions for the molar enthalpy for the seven species that appear in the complete combustion reactions of Ar, $CH_4$, $CO_2$, $H_2$, $H_2O$, $N_2$, and $O_2$, respectively. The profiles presented in these figures are for the non-dimensional quantity described by (2), with coefficients based on GRI-Mech 3.0 for one profile and based on NASA (1993 edition) values for the other. The total range of temperature associated with the polynomial coefficients from $T_{min}$ to $T_{max}$ is mentioned, for both the GRI-Mech 3.0 and the 1993 NASA coefficients. For these seven species, the intermediate temperature ($T_{mid}$) separating the low- and high-temperature subranges is the same for both sets of polynomial coefficients, being 1,000 K. The figures show the smooth transition at $T_{mid}$.

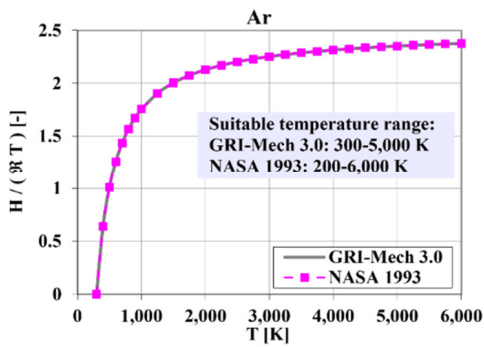

Fig. 1. Normalized molar enthalpy of Ar versus temperature as described by two 6-term polynomials.

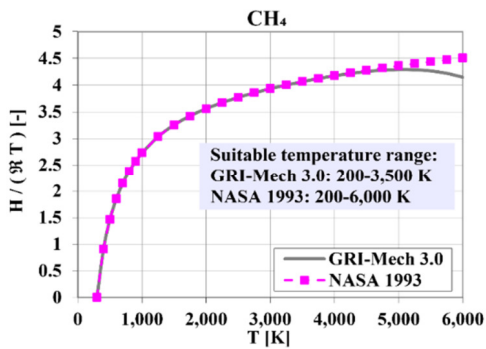

Fig. 2. Normalized molar enthalpy of $CH_4$ versus temperature as described by two 6-term polynomials.

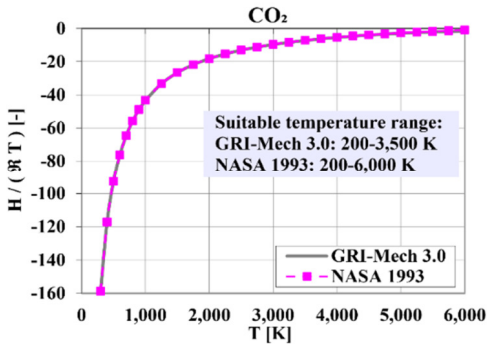

Fig. 3. Normalized molar enthalpy of $CO_2$ versus temperature as described by two 6-term polynomials.

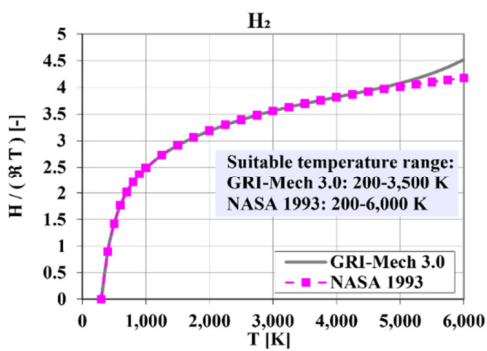

Fig. 4. Normalized molar enthalpy of $H_2$ versus temperature as described by two 6-term polynomials.

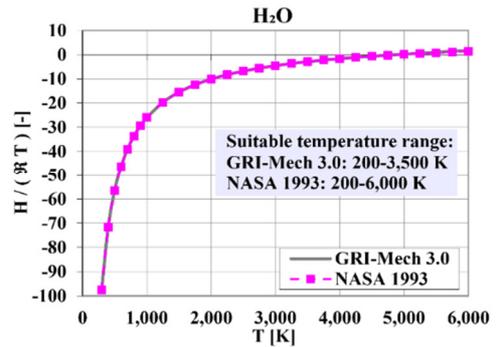

Fig. 5. Normalized molar enthalpy of $H_2O$ versus temperature as described by two 6-term polynomials.

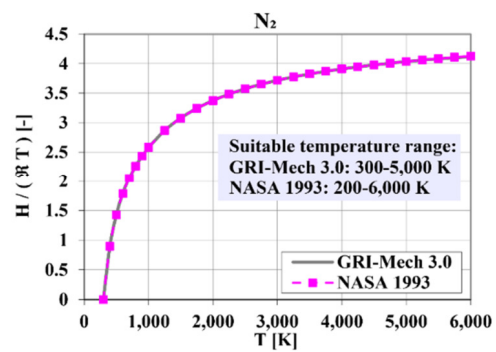

Fig. 6. Normalized molar enthalpy of $N_2$ versus temperature as described by two 6-term polynomials.

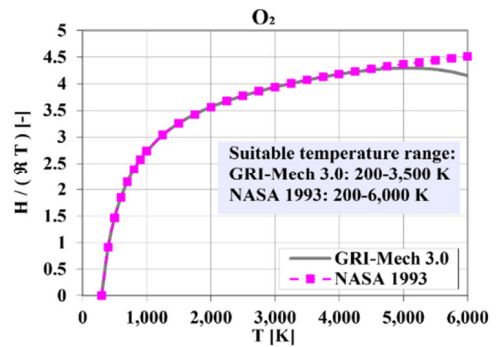

Fig. 7. Normalized molar enthalpy of $O_2$ versus temperature as described by two 6-term polynomials.

The NASA coefficients for these seven species were designed for a wider temperature range with an upper-temperature limit of 6,000 K, while the upper-temperature limit for GRI-Mech 3.0 is 3,500 K for five species and 5,000 K for Ar and $N_2$. Such a very hot temperature allowed by the NASA coefficients is not common in terrestrial combustion applications, and thus the GRI-Mech 3.0 narrower temperature range is generally reasonable. The two profiles are in good agreement for Ar, $CO_2$, $H_2O$, and $N_2$. No noticeable deviation occurs for $CH_4$, $H_2$, and $O_2$ before approximately 5,000 K. Since the maximum AFT predicted using GRI-Mech 3.0 for oxy-methane with assumed complete combustion was 5,153.6 K, which is not far from 5,000 K, the use of GRI-Mech 3.0 was considered acceptable despite being used outside the exact temperature range of its polynomial coefficients in some cases.

Figure 8 shows how the dimensional molar enthalpy for the seven selected species varied with the absolute temperature according to the polynomial coefficients of GRI-Mech 3.0. To convert from a non-dimensional version of the molar enthalpy to the dimensional version, the value of 8.314510 J/mol.K was used for the universal gas constant, $\Re$.

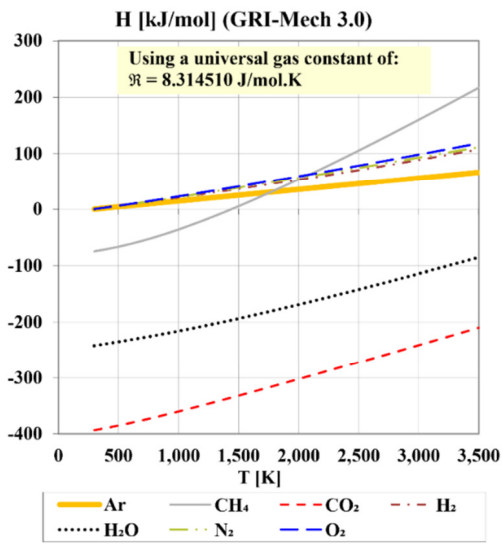

Fig. 8.     Molar enthalpy (based on GRI-Mech 3.0 coefficients) of the 7 species that appear in the single-step complete combustion reaction of CH$_4$ (in air or oxygen) or H$_2$ (in air or oxygen).

The used value of the universal gas constant was cited by NASA in its 1993 edition of the 7-coefficient fitting functions for thermodynamic properties and is the same value published later with an extended version (NASA 9-coefficient polynomial fitting functions) in 2002, where the non-dimensional molar enthalpy was also allowed to depend on $1/T^2$ and $\ln(T)/T$ [38], and thus two terms were added to the fitting function. CEA used this newer 9-coefficient fitting structure, but GRI-Mech 3.0 uses the original 7-coefficient polynomial structure. This is justifiable, because the increased complexity in the newer fitting function may not necessarily increase the fitting accuracy within the temperature range of interest for GRI-Mech 3.0 applications, while the newer fitting functions enabled very large temperatures up to 20,000 K.

The CEARUN values were considered the principal, more accurate values, while the GRI-Mech 3.0 values were considered supplementary benchmarking values for comparison. Table I lists the computed AFT for oxy-methane complete combustion. The CEARUN value 5,166.47 K was very close to the GRI-MECH 3.0 prediction using the MS Excel Goal Seek tool (5,153.68 K), and the relative deviation was only 0.248%. Table II shows that the oxy-hydrogen complete combustion AFT from CEARUN was 4,930.56 K, which is 235.91 K lower than that of methane, and GRI-MECH 3.0 provided a lower prediction by 40.35 K.

TABLE I.     AFT FOR OXY-METHANE COMBUSTION

| Single-Step Complete Combustion | CEARUN | GRI-Mech 3.0 (Excel) |
|---|---|---|
| AFT | 5,166.47 K | 5,153.68 K |
| Absolute difference | - | 12.79 K |
| Absolute difference as a percentage of CEARUN value | - | 0.248% |

TABLE II.     AFT FOR OXY-HYDROGEN COMBUSTION

| Single-Step Complete Combustion | CEARUN | GRI-Mech 3.0 (Excel) |
|---|---|---|
| AFT | 4,930.56 K | 4,890.21 K |
| Absolute difference | - | 40.35 K |
| Absolute difference as a percentage of CEARUN value | - | 0.818% |

As shown in Tables III and IV, changing the oxidizer from pure oxygen to air significantly reduces AFT for both fuels when assuming complete combustion. The reduction was 2,840.12 K for methane (55.0% AFT in oxy-fuel complete combustion), and 2,410.23 K for hydrogen (48.9% AFT in oxy-fuel complete combustion). With air as the oxidizer, the complete combustion of hydrogen is hotter than that of methane. This is opposite to the situation of oxy-fuel complete combustion, with the difference in the estimated AFT being 193.98 K. The relative deviation between GRI-Mech 3.0 and CEARUN is small, less than 0.2%, for either fuel.

TABLE III.     AFT FOR AIR-METHANE COMBUSTION

| Single-Step Complete Combustion | CEARUN | GRI-Mech 3.0 (Excel) |
|---|---|---|
| AFT | 2,326.35 K | 2,330.55 K |
| Absolute difference | - | 4.20 K |
| Absolute difference as a percentage of CEARUN value | - | 0.181% |

TABLE IV.     AFT FOR AIR-HYDROGEN COMBUSTION

| Single-Step Complete Combustion | CEARUN | GRI-Mech 3.0 (Excel) |
|---|---|---|
| AFT | 2,520.33 K | 2,524.36 K |
| Absolute difference | - | 4.03 K |
| Absolute difference as a percentage of CEARUN value | - | 0.160% |

Investigating the AFTs in Tables V-VIII, when chemical equilibrium is used as the criterion to find the composition of the combustion products rather than enforcing a specific composition with a few species, reveals the impact of using a detailed realistic reaction mechanism or allowing for more species to appear through chemical equilibrium computation. The AFT of oxy-fuel and air-fuel combustion of methane decreased by 2,116.35 K and 102.10 K, which is 41.1% and 4.39% of the respective complete combustion values, respectively. The AFT of oxy-fuel and air-fuel combustion of hydrogen decreased by 1,856.05 K and 141.71 K, which is 37.6% and 5.62% of the respective complete combustion values, respectively. For both fuels, the effect of applying equilibrium computation is relaxed in air-fuel combustion, where the AFT is already much lower than its higher value with oxy-fuel combustion. As an additional remark, the relative deviation between the AFT predictions of CEARUN and the Cantera implementation of GRI-Mech 3.0 for oxy-fuel combustion is much smaller, less than 0.08% for methane or hydrogen, than for air-fuel combustion (3.06% for methane and 3.50% for hydrogen).

TABLE V.    AFT FOR OXY-METHANE COMBUSTION

| Chemical Equilibrium Combustion | CEARUN | GRI-Mech 3.0 (Cantera) |
|---|---|---|
| AFT | 3,050.12 K | 3,052.06 K |
| Absolute difference | - | 1.94 K |
| Absolute difference as a percentage of CEARUN value | - | 0.064%[a] |

a. Smallest magnitude of the relative deviation among all 8 cases.

TABLE VI.    AFT FOR OXY-HYDROGEN COMBUSTION

| Chemical Equilibrium Combustion | CEARUN | GRI-Mech 3.0 (Cantera) |
|---|---|---|
| AFT | 3,074.51 K | 3,076.92 K |
| Absolute difference | - | 2.41 K |
| Absolute difference as a percentage of CEARUN value | - | 0.078% |

TABLE VII.    AFT FOR AIR-METHANE COMBUSTION

| Chemical Equilibrium Combustion | CEARUN | GRI-Mech 3.0 (Cantera) |
|---|---|---|
| AFT | 2,224.25 K | 2,156.25 K |
| Absolute difference | - | 68.00 K |
| Absolute difference as a percentage of CEARUN value | - | 3.057% |

TABLE VIII.    AFT FOR AIR-HYDROGEN COMBUSTION

| Chemical Equilibrium Combustion | CEARUN | GRI-Mech 3.0 (Cantera) |
|---|---|---|
| AFT | 2,378.62 K | 2,295.29 K |
| Absolute difference | - | 83.33 |
| Absolute difference as a percentage of CEARUN value | - | 3.503%[a] |

a. Largest magnitude of the relative deviation among all 8 cases.

## IV.    CONCLUSIONS

This study calculated 16 AFT values for stoichiometric combustion for different methane and hydrogen combustion conditions. Reactants were specified so that no excess fuel or oxidizer exists beyond what is theoretically needed for complete combustion. The initial temperature was 298.15 K as a common reference value and the absolute pressures were kept constant at a common reference value of 1 atm. Half of the total cases, 8 out of 16, corresponded to methane gas, and the other half corresponded to molecular hydrogen gas. For each fuel, half of the cases, 4 out of 8 per fuel, corresponded to the use of pure oxygen as oxidizer, and the other half corresponded to the use of a mixture that mimicked dry air. For each oxidizer type, two ways were used to specify the final composition after combustion, namely, by enforcing a complete combustion, the simplest modeling method, or solving a chemical equilibrium problem that requires more computations. Finally, for each way of specifying the final mixture, one case was modeled using the web-based tool CEARUN by NASA, while another supplementary case was modeled using the GRI-Mech 3.0 reaction mechanism either through MS Excel or through the Cantera Python add-on package. The results show that oxy-combustion is capable of reaching much higher temperatures than air-combustion, since the difference was 825.87 K in the case of methane and 695.89 K in the case of hydrogen. Furthermore, simplified global reaction mechanisms can give a reasonable estimation of AFT in air-fuel combustion when compared with detailed mechanisms, but they are largely inaccurate for oxy-fuel combustion. The study also showed that the Cantera package with the embedded GRI-Mech 3.0 is relatively accurate for estimating AFT, as it gave values similar to those obtained by the more comprehensive tool CEARUN.

This study can be useful in the broad fields of combustion modeling, numerical simulation, and curve-fitting of thermodynamic properties through the following:

- Quantitatively identify the AFTs of two specific fuels with conventional air-combustion and non-conventional oxy-combustion, under clearly described conditions. The modeled cases are reproducible for verification or use as test cases for an independent modeling tool, and for problems that need a good estimate of combustion peak temperatures.

- Provided examples that showed the impact of simplification in modeling AFT calculations.

- Proposed a minor correction in one of the GRI-Mech 3.0 coefficients (for the nitrogen diatomic gaseous species) providing details and justification.

- Presented a validation analysis for Cantera as a tool for computing AFT through the reported relative deviation when compared to the more-detailed NASA-developed CEARUN tool for methane and hydrogen.

- Can be regarded as an overview of the valuable and free simulation tools of CEARUN and Cantera, as well as the GRI-Mech 3.0 implementation of the NASA 7-coefficient polynomials, and the difference between this 7-coeffecient version of NASA fitting functions for thermodynamic properties and the newer 9-coefficient version.

- The visual comparison between of the results after using the NASA 1993 and the GRI-Mech 3.0 coefficients to predict the normalized molar enthalpy for different species commonly used in combustion is also a useful component of the current study. In [39], the AFT of hydrogen was mentioned as 2,382 K. The current study showed a much larger AFT when oxygen was the oxidizer (3,074.51 K). The cited value was close to the value found here for air-hydrogen combustion with chemical equilibrium (2,378.62 K). Moreover, the cited value was not reproducible, given the lack of knowing the exact initial conditions (reactants temperatures), unless a guess of reasonable values is provided. In the current study, such details are provided.